# Investigation of Space-Time Structure of the Discharge with an Electrolytic Anode and Face-Type, Air Half-Space Directed Cathode (Gatchina's Discharge)

Emelin S.E., Astafiev A.M., Pirozerski A.L.

*Faculty of Physics, Saint-Petersburg State University, Russia*

Despite of its limited capabilities for ball-lightning modeling, the chemically active dust plasma of electric discharges is a very interesting physical object. Gatchina's discharge [1], which gives a possibility to study the dust-gas fireball, represents a complex non-stationary process, combining the creation and the destruction of relatively long-living high-enthalpy microscopic states with non-ideal dust plasma and with a gas-dynamic form, resulting in appearance of a spherical luminous object [2-4]. In the present work we studied spatial structure of the discharge and its dynamics with the help of electric and optical measurements.

Discharge installation consisted of the capacitor store in capacity up to 4 mF to be charged up to 5 kV and connected to Gatchina's discharger through an additional air gap and a high-voltage winding of the pulse ignition transformer with inductance 7.6 mH. The discharger represented a cylindrical glass of diameter 17 cm and of height 21 cm, filled to 20 cm by a weak solution of sodium hydrocarbonate or nitric acid. The wire ring anode of diameter 15 cm settled down at the bottom of the glass. Coal or iron cylindrical cathode of diameter 6-8 mm, grounded, established on the end of a long steel rod, was isolated from water by a quartz tube (inner diameter 9 mm, outer –12 mm) and fixed vertically in an aperture at the bottom of the glass.

Discharge current and anode voltage dependences on the time (Fig. 1, left) reveal two prominent features: increase of the resistance at the current reduction and a break of the voltage-current characteristic, observed in the chosen example at 80-th millisecond.

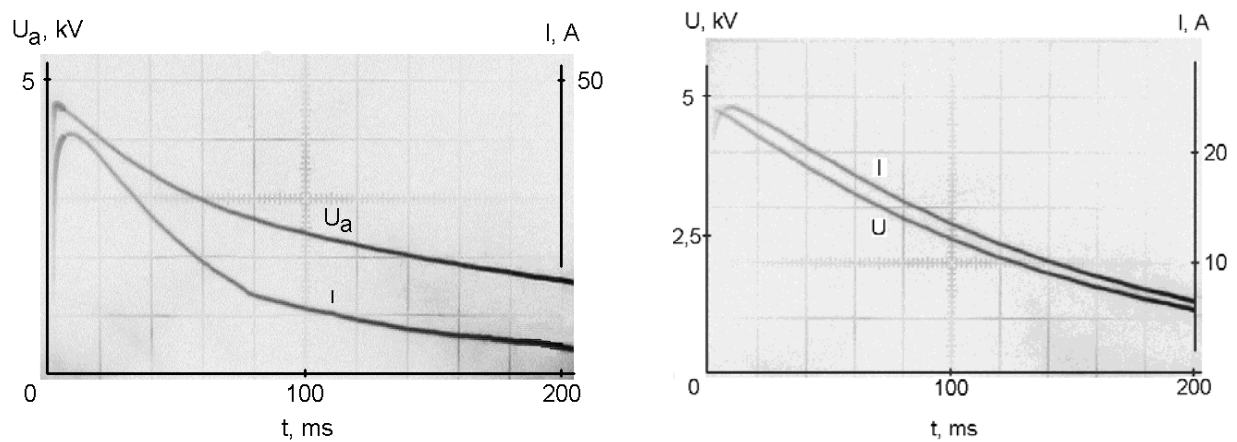

Fig. 1. Left - oscillogram of the discharge current I and anode voltage $U_a$, right - the same for discharge between two ring wire electrodes placed under water.

At the same time similar dependence for the discharge between two ring wire electrodes, located under water, (Fig. 1 right) testifies to an insignificant deviation of conductivity of the electrode-solution system within a characteristic range of parameters. It means that the discharge and area of a solution near the discharge exert strong influence upon resistance of the electric circuit.

To find a voltage drop at this part of the electric circuit two probes which were settled down on small depth close (1 cm) and far (7 cm) from the cathode were used. The characteristic oscillogram of probe signals is given at Fig. 2. One can see that the voltage drop at the discharge does not exceed 2 kV and near the cathode at a maximum of a current - 0,6 kV. The break at 80-th millisecond is present only at the signal of the remote probe, coinciding with its maximum and the maximal difference from the signal of the near probe, in which the break takes place at 50-th millisecond. The minimal distinction of probe signals is observed from 20-th up to 50-th millisecond, i.e. after some accumulation of discharge products and also near the current breakdown.

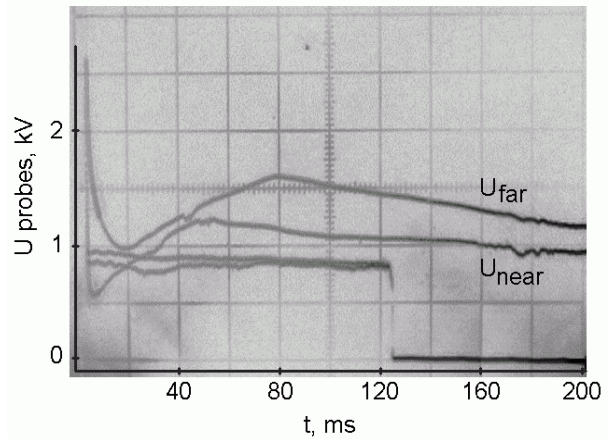

Fig. 2. The oscillogram of probe signals. The beginning of the second trace corresponds to 240 ms.

At Fig. 3 on the left the oscillogram of relative intensity of optical radiation and the discharge current is presented and, on the right, - the discharge with compulsory sharp breakdown of the current at 80-th millisecond.

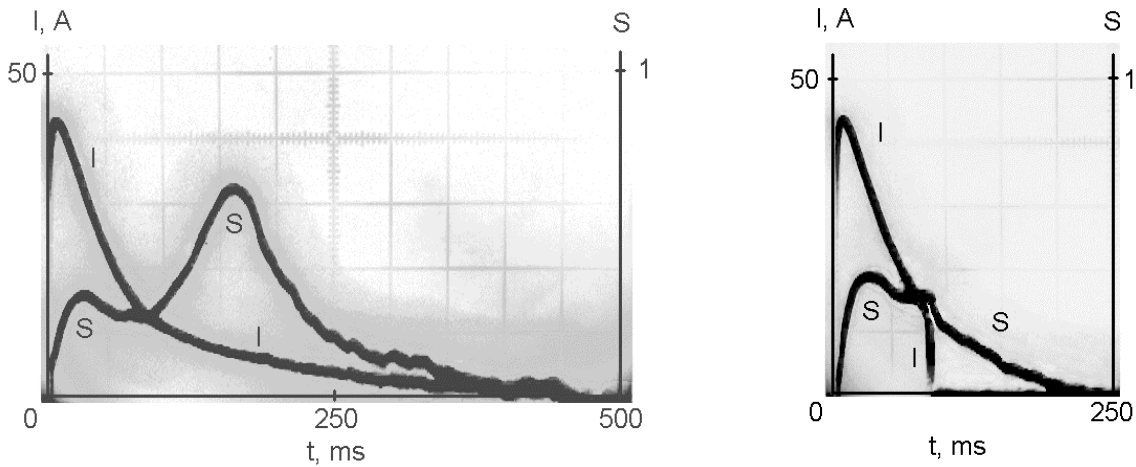

Fig. 3. Optical radiation intensity and discharge current for a solution of sodium hydrocarbonate. On the right - the discharge with compulsory sharp breakdown of the current.

Optical radiation of the discharge starts to intensify to 50-th millisecond and has the greatest intensity at repeatedly reduced current. When the current breakdown occurs in the beginning of increasing of the radiation the course of relaxation is quickly restored and continues the course characteristic for afterglow.

Spatial distribution of emission spectra of the discharge and afterglow has been studying with use of solutions and of materials of the cathode with essentially different spectra. As result it was found out that when the maximal current is less than 50 A the material of the cathode concentrates in the central area of the discharge up to 80-th millisecond, and the substance dissolved in water is shown at the discharge very poorly. At 80-th millisecond a strong radiation of atoms of the dissolved substance (Na) appears and fills in all area of afterglow.

This dynamics can be clarified with the help of a sequence of the color video frames made with an interval 20 ms and presented at Fig. 4.

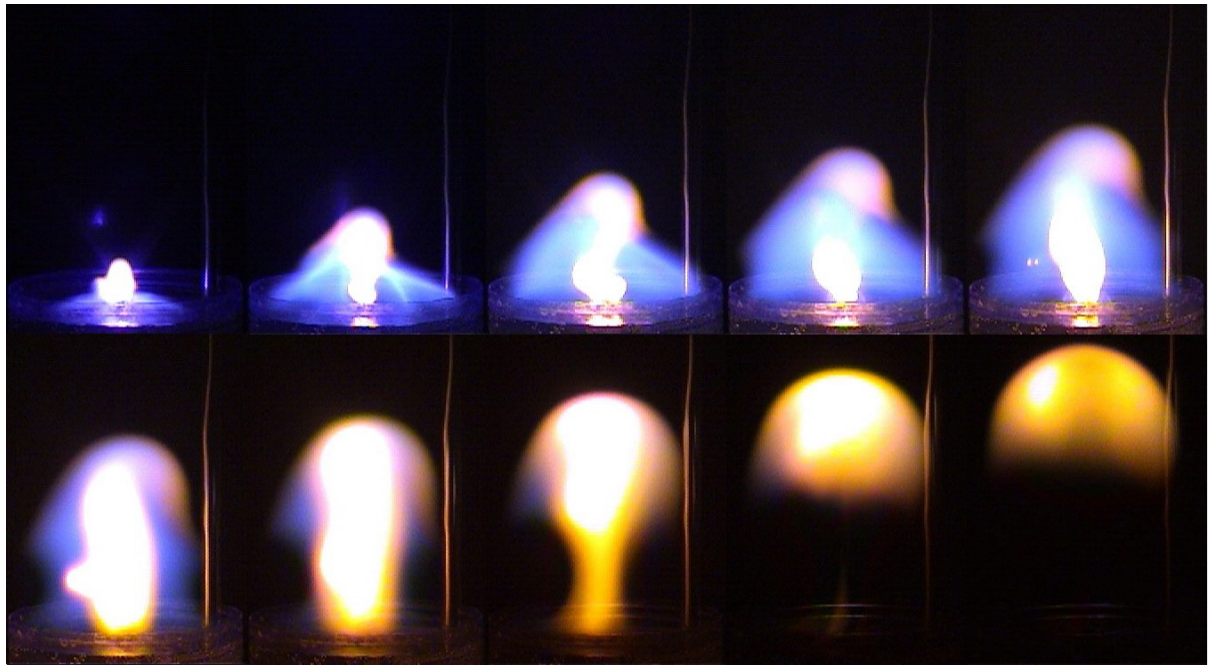

Fig. 4. Videoshooting of the discharge and the afterglow.

In this case the sodium hydrocarbonate solution and the coal electrode were used, the current breakdown has taken place at the 7-th frame. Accumulation of the water evaporated within anode layer with the discharge distributed within it occurs during the time corresponding to the upper row of images. At 6-th and 7-th frames the basis of the central erosive jet is expanded on water, jet involves products of the space discharge further forming a fireball.

For reduction of the content of solids in the discharge the variant with two electrolytic electrodes on the basis of a nitric acid solution was considered. Received objects looked like a vaporous cloud and had smaller optical radiation, spectrum of which contained weak lines of casual impurity (Fig.5).

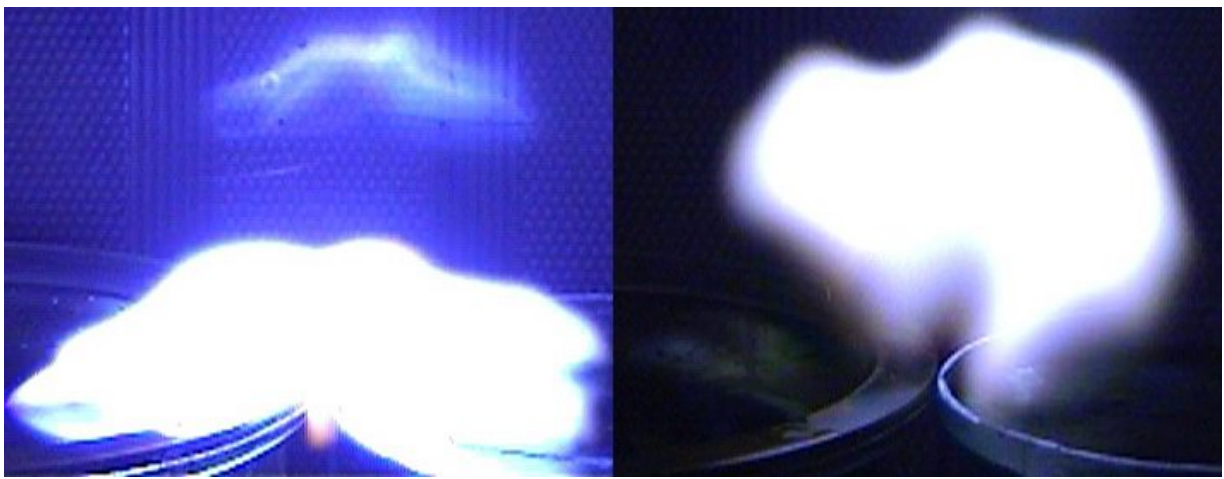

Fig. 5. The discharge with two electrolytic electrodes on the basis of the nitric acid.

However the discharge with the central iron electrode gives rise to objects of the spherical form even at small maximal currents when the vortex and erosion of the cathode are weak. On Fig. 6 the images of a fireball after separation from the discharge (170 ms) and a sequence of spectra prior to the beginning of the object separation (120 ms) are presented. At the orange-yellow area (on the left) a molecular spectrum of FeO and a doublet of the sodium are visible, other lines belong to the atomic spectrum of the iron. Appearance of the doublet of sodium about the cathode at 5-th frame indicates localization of the discharge.

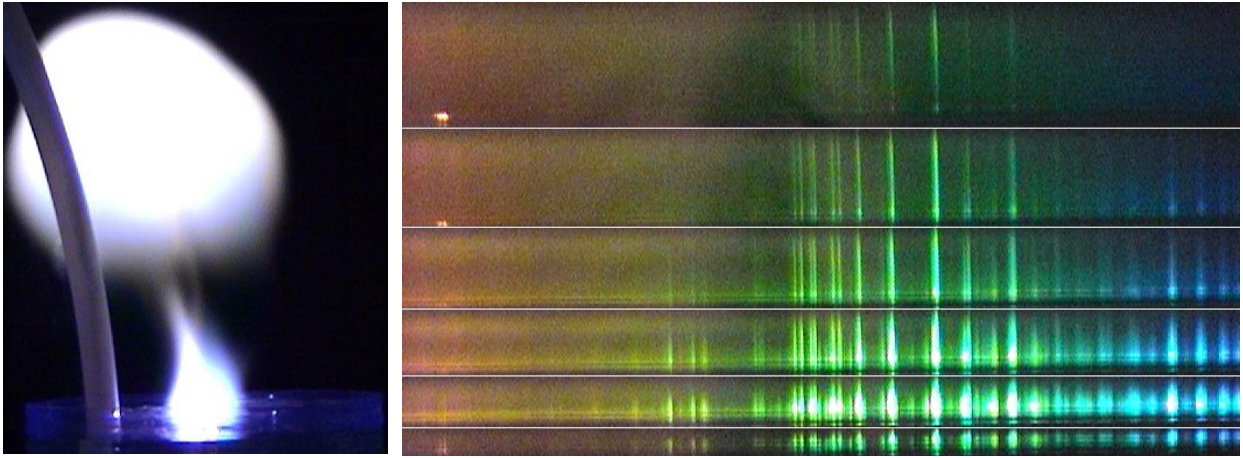

Fig. 6. At the left - discharge with the nitric acid and central iron cathode, at the right - its spectra following bottom-up with interval 20 ms.

Results of the experiment allow conclude that during the large current the discharge is distributed above electrolytic anode and effectively creates chemically active substance, but evaporation of water occurs without a drop phase. During a small current the discharge is located about the cathode, resistance of the solution concentrates near an anode spot, and erosion of the electrolytic anode is controlled by the drop phase actively bringing dissolved substance into volume of chemically active plasma. Catalyzing action of the dust components formed on the base of the cathode substance, of uncontrolled dielectric impurities from some discharger's parts and of the dissolved substance results in appearance of a nonideal dust chemical plasma fireball.